\documentclass[conference, letterpaper]{IEEEtran}
\IEEEoverridecommandlockouts
\usepackage{cite}
\usepackage{amsmath,amssymb,amsfonts}
\usepackage{algorithm,algorithmic}
\usepackage{graphicx}
\usepackage{textcomp}
\usepackage{amsmath}  
\usepackage{amsfonts}  
\usepackage{amssymb}  
\usepackage{soul}
\usepackage{enumitem}
\usepackage{xcolor}
\usepackage{booktabs}
\usepackage{cuted}
\usepackage{svg}
\usepackage{tcolorbox}
\usepackage{bm}
\usepackage[letterpaper, right=0.622in, left=0.622in, top=0.73in, bottom=1.02in]{geometry}

\def\BibTeX{{\rm B\kern-.05em{\sc i\kern-.025em b}\kern-.08em
    T\kern-.1667em\lower.7ex\hbox{E}\kern-.125emX}}

\makeatletter
\def\@fnsymbol#1{\ensuremath{\ifcase#1\or *\or \dagger\or \ddagger\or \mathsection\or \mathparagraph\or \|\or **\or \dagger\dagger\or \ddagger\ddagger \fi}}
\makeatother

\begin{document}

\title{Sum Rate Maximization for Movable Antenna-Aided Downlink RSMA Systems}
\author{

\IEEEauthorblockN{Cixiao Zhang\IEEEauthorrefmark{2}, Size Peng\IEEEauthorrefmark{2}, Yin Xu\IEEEauthorrefmark{2}, Qingqing Wu\IEEEauthorrefmark{3}, Xiaowu Ou\IEEEauthorrefmark{2}, Xinghao Guo\IEEEauthorrefmark{2}, Dazhi He\IEEEauthorrefmark{2} and Wenjun Zhang\IEEEauthorrefmark{2}}
\IEEEauthorblockA{\IEEEauthorrefmark{2}Cooperative Medianet Innovation Center (CMIC), Shanghai Jiao Tong University, Shanghai 200240, China.\\
\IEEEauthorrefmark{3}Department of Electronic Engineering, Shanghai Jiao Tong University, Shanghai, China.\\
Email: \{cixiaozhang, sjtu2019psz, xuyin, qingqingwu, xiaowu\_ou, guoxinghao, hedazhi, zhangwenjun\}@sjtu.edu.cn
}


    \thanks{

    The corresponding author is Yin Xu (e-mail: xuyin@sjtu.edu.cn).
    
}
}

\maketitle

\begin{abstract}
Rate splitting multiple access (RSMA) is regarded as a crucial and powerful physical layer (PHY) paradigm for next-generation communication systems. Particularly, users employ successive interference cancellation (SIC) to decode part of the interference while treating the remainder as noise. However, conventional RSMA systems rely on fixed-position antenna arrays, limiting their ability to fully exploit spatial diversity. This constraint reduces beamforming gain and significantly impairs RSMA performance. To address this problem, we propose a movable antenna (MA)-aided RSMA scheme that allows the antennas at the base station (BS) to dynamically adjust their positions. Our objective is to maximize the system sum rate of common and private messages by jointly optimizing the MA positions, beamforming matrix, and common rate allocation. To tackle the formulated non-convex problem, we apply fractional programming (FP) and develop an efficient two-stage, coarse-to-fine-grained searching (CFGS) algorithm to obtain high-quality solutions. Numerical results demonstrate that, with optimized antenna adjustments, the MA-enabled system achieves substantial performance and reliability improvements in RSMA over fixed-position antenna setups.

\end{abstract}

\section{Introduction}
Multiple-input multiple-output (MIMO) technology, extensively studied and foundational to next-generation communication systems \cite{MIMOSurvey}, relies on beamforming as a key technique. Beamforming, a critical technology in MIMO systems, leverages spatial domain freedom to achieve precise angular selectivity, effectively compensating for significant path loss. This spatial manipulation enhances signal quality and system performance by directing energy toward desired directions and minimizing interference. However, traditional antenna arrays are fixed in position, limiting their adaptability to dynamic channel conditions. As a result, they may struggle to maintain optimal performance in rapidly changing environments, underscoring the need for more flexible solutions.

To overcome the limitations of fixed-position antennas (FPAs) and fully utilize spatial degrees of freedom (DoFs), the movable antenna (MA) and the fluid antenna (FA) have been introduced \cite{MASurvey1, FASurvey1, FA_SURVEY_RSMA, FALEE}. By dynamically adjusting the antenna positions within the feasible regions to change the phases across various propagation paths, MAs and FAs can enhance the performance of wireless communication. Specifically, in \cite{FAISAC}, a base station (BS) served both a communication user and a sensing target, with fluid antennas equipped at both the BS and the communication user. This system enhanced the downlink communication rate by meeting sensing beam pattern gain requirements. As presented in \cite{Peng2024JointAP}, the authors investigated a full-duplex integrated sensing and communication (FD-ISAC) system assisted by MAs. They verified that an MA-aided flexible beamforming reduced self-interference and enhanced sensing and communication performance. As reported in \cite{zhu2024suppressing}, MAs sufficiently alleviated beam squint under a wideband multiple-input-single-output (MISO) scenario. In \cite{MA_PAN}, an MA-aided ISAC system was enhanced with a reconfigurable intelligent surface (RIS) to improve communication and sensing performance in dead zones. In \cite{MARELATED1}, the authors studied the utility of movable antennas in multiple-input single-output (MISO) interference channels, proposing an iterative algorithm to optimize MA positions and beamforming, enhancing performance and reducing transmitter complexity.

On the other hand, rate splitting multiple access (RSMA) has emerged as a flexible and effective PHY-layer transmission paradigm for non-orthogonal transmission, interference management, and multiple access strategies envisioned for 6G \cite{RSMA_survey1}. By fully leveraging rate splitting (RS), beamforming, and successive interference cancellation (SIC), RSMA enables partial interference decoding while treating the remaining interference as noise.\cite{RSMA_Related1} proposed a robust beamforming and rate optimization algorithm to enhance spectral efficiency and transmission robustness in RIS-aided symbiotic radio systems using RSMA. In \cite{RSMA_HAIXIA}, a full-duplex cooperative rate-splitting scheme was proposed for downlink multicast, where cell-center users assisted cell-edge users via distributed beamforming and power-splitting for energy efficiency. \cite{FIRSTRSMAISAC} was the first work to propose an RSMA-based DFRC architecture. This approach maximized communication rates and improved radar sensing, achieving superior performance over SDMA-based DFRC without requiring additional radar sequences, simplifying the architecture, and enhancing system performance. However, a critical challenge remains, particularly in overloaded scenarios, where the limited beamforming gains from conventional fixed-position antenna arrays restrict RSMA's performance potential.

Inspired by these promising methods, we are motivated to investigate effective configuration strategies for MAs to fully leverage the spatial DoFs and enhance the RSMA system performance. Notably, no prior work has examined the integration of RSMA with MAs. To fill this gap, our work focuses on a downlink MA-aided  RSMA system to maximize the sum rate. Our objective is to maximize the users’ sum rate by jointly optimizing beamforming matrices, splitting rates, and the positions of MAs. We employ an Alternating Optimization (AO) method to handle the coupled parameters. Specifically, to solve this non-convex problem, we use the Fractional Programming (FP) method from \cite{FPmethod} to optimize the beamforming matrix and adopt a coarse-to-fine-grained searching (CFGS) method to identify high-quality sub-optimal antenna positions. Our contributions are summarized as follows.

\begin{itemize}
    \item To the best of our knowledge, this is the first study to examine the performance of MAs within an RSMA system. We formulate an optimization problem to maximize the sum rates of all users by jointly optimizing the beamforming matrix, common rate splitting, and antenna positions.
    
    \item An FP framework is utilized to reformulate the objective function. We apply an AO method to handle the coupled optimization parameters, decomposing the problem into four sub-problems. Specifically, we propose an efficient algorithm based on a CFGS approach to update the antenna positions, leading to substantial performance improvements.

    \item Numerical results demonstrate that the proposed MA-aided RSMA system outperforms the FPA configuration, achieving superior performance with our algorithm. Furthermore, our algorithm exhibits robustness across varying power budgets and user antenna configurations.
    
\end{itemize}

\textit{Notations:} $\mathbf{x}(n)$, $\mathbf{x}^{T}$, $\mathbf{x}^{*}$, $\text{Tr}(\mathbf{X})$and $(\mathbf{X})^{-1}$  denote the $n^{th}$ entry of $\mathbf{x}$, the transpose of $\mathbf{x}$, the conjugate of $\mathbf{x}$, the trace of $\mathbf{X}$ and the inverse of $\mathbf{X}$, respectively.

\begin{figure}[t]
    \centering
    \includegraphics[width=0.46\textwidth]{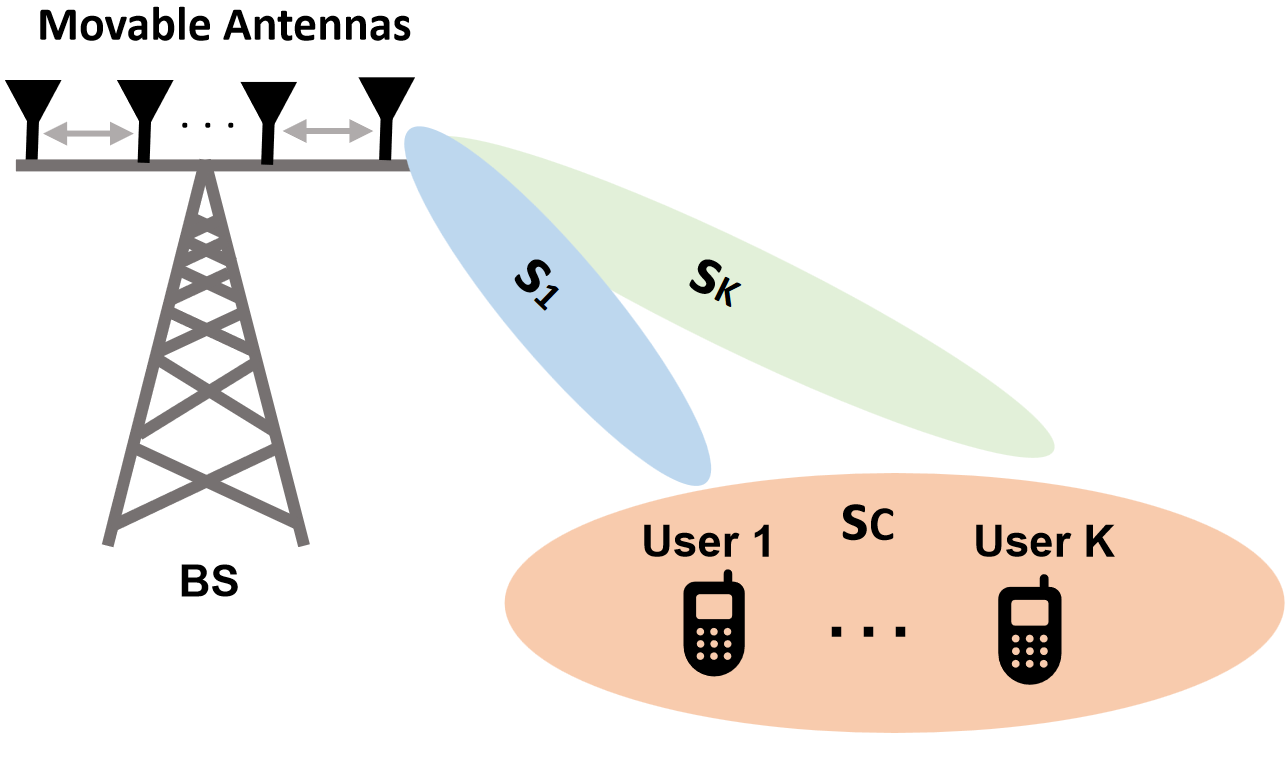}
    \caption{System model of an MA-aided RSMA system.}
    \vspace{-0.5cm}
    \label{fig: system model}
\end{figure}

\section{System Model}
\begin{figure}[t]
    \centering
    \includegraphics[width=0.48\textwidth]{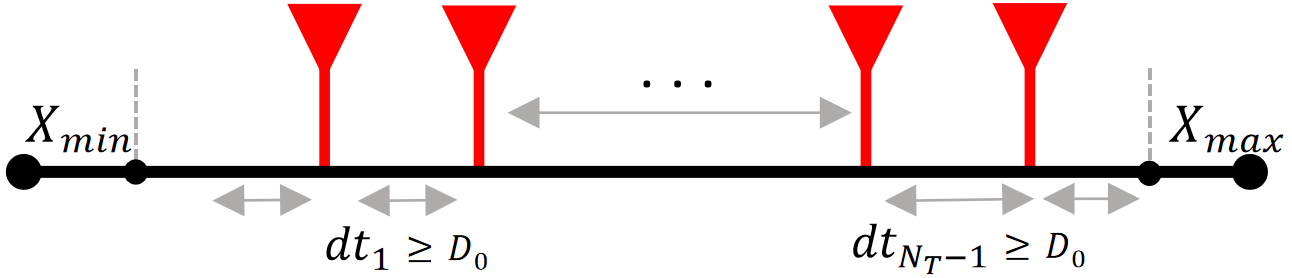}
    \caption{Linear movable antenna model.}
    \vspace{-0.5cm}
    \label{fig: antenna spacing}
\end{figure}

In this paper, we consider a widely adopted RSMA scheme known as 1-layer RSMA \cite{RSMA_survey1, RSMA_survey2}, involving a base station (BS) with $N_T$ linear MAs serving $K$ single-antenna users, as illustrated in Fig. \ref{fig: system model}. The users are indexed by $\mathcal{K}=\{1,2,\cdots, K\}$. For simplicity, we consider a 1-dimension array. The antenna positions can be flexibly adjusted within the linear region $[X_{\min}, X_{\max}]$, as illustrated in Fig. \ref{fig: antenna spacing}.

\subsection{Channel Model}
Denote $\mathcal{N}=\{1,2,\cdots, N_T\}$ as the set of MAs and the positions of them are denoted as a vector $\mathbf{x}=[x_1,..., x_{N_T}]^T$. Assume that the channel from the BS to each user experiences $L_p$ propagation paths. According to \cite{MA_channel_model}, the field response vector (FRV) of the far-field channel is 
\begin{equation}
\scalebox{0.9}{$
\mathbf{a}_{k}(x_i) = \begin{bmatrix} e^{j\frac{2\pi }{\lambda} x_i\cos(\theta_{k,1})},\cdots, e^{j\frac{2\pi }{\lambda} x_i\cos(\theta_{k,L_p})} \end{bmatrix}^T\in \mathbb{C}^{L_p\times 1},$
}
\end{equation}
where $\theta_{k,l}$ represents the angle of departure (AOD) of the $l^{th}$ path for the $k^{th}$ user and $\lambda$ denotes the wavelength. By combining all the FRVs, we obtain the field response matrix (FRM) as
\begin{equation}
    \mathbf{A}(\mathbf{x})=[\mathbf{a}_{k}(x_1),\cdots,\mathbf{a}_{k}(x_{N_T})]\in \mathbb{C}^{L_p\times N_T}.
\end{equation}

Next, we define $\mathbf{1} \in \{1\}^{L_p\times1}$ as the all-one FRV. The path response matrix (PRM) between the BS and the $k^{th}$ user as $\boldsymbol{\Sigma}_k=\text{diag}([\rho_{k,1},...,\rho_{k, L_p}]^T) \in \mathbb{C}^{L_p\times L_p}$, where $\rho_{k,l}$ denotes the gain of the $l^{th}$ path for the $k^{th}$ user. Consequently, the communication channel between the BS and $k^{th}$ user is expressed as
\begin{equation}
    \mathbf{h}_k = \mathbf{A}^H(\mathbf{x})\boldsymbol{\Sigma}_k \mathbf{1}\in \mathbb{C}^{N_T\times 1}.
\end{equation}

\subsection{Signal Model}
In a 1-layer RSMA, each user message $W_k$ is divided into a common sub-message $W_{c,k}$ and a private sub-message $W_{p,k}$. The private sub-messages are encoded into $\{s_1,s_2,...,s_K\}$ respectively, while the common sub-messages are combined and encoded into a single common stream $s_{c}$. 
It is assumed that each stream in $\{s_1,s_2,...,s_K,s_{c}\}$ satisfies $\mathbb{E}[s_is_i^H]=1$, and $\mathbb{E}[s_is_j^H]=0,\forall i \neq j.$ The beamforming vector for $k^{th}$ user's private stream is denoted as $\mathbf{f}_k$, and the beamforming vector for the common stream is $\mathbf{f}_{c}$. Consequently, the beamforming matrix is represented as
\begin{equation}
    \mathbf{F}=[\mathbf{f}_1,...,\mathbf{f}_K,\mathbf{f}_{c}]\in \mathbb{C}^{N_T \times (K+1)}.
\end{equation}
Accordingly, the baseband received signal at the $k^{th}$ user is given by
\begin{equation}
    y_k = \mathbf{h}_k^H(x)\mathbf{f}_ks_k + \sum\limits_{j=1, j\neq k}^{K+1}\mathbf{h}_k^H \mathbf{f}_j s_j + n_k, 
\end{equation}
\begin{figure*}[t]
\begin{align}
\scalebox{0.92}{
$\hat{\mathcal{G}}(\mathbf{F}, \mathbf{r_c}, \mathbf{x}, \boldsymbol{\alpha},\boldsymbol{\beta}) = \underbrace{\frac{1}{\log(2)}\sum\limits_{k=1}^{K}[\log(1+\alpha_k)-\alpha_k]+\sum\limits_{k=1}^{K}r_{c,k}}_{\hat{\mathcal{G}}_0} + \frac{2}{\log(2)}\sum\limits_{k=1}^{K}\underbrace{\sqrt{\alpha_k+1}\Re\big(\beta_k^*\mathbf{h}_k^H(\mathbf{x})\mathbf{f}_k\big)}_{\hat{\mathcal{G}}_{1,k}} - \frac{1}{\log(2)}\sum\limits_{k=1}^{K}\underbrace{|\beta_k|^2( \sum\limits_{j=1}^{K}|\mathbf{h}^H_k(\mathbf{x})\mathbf{f}_j|^2+\sigma_k^2)}_{\hat{\mathcal{G}}_{2,k}}$},
\label{convexG}
\end{align}
\begin{align}
\scalebox{0.98}{$
t_k(\mathbf{F},\mu_k,\eta_k)=\log(1+\mu_k)-\mu_k+2\sqrt{\mu_k+1}\Re(\eta_k^*\mathbf{h}_k^H(x)\mathbf{f}_c)-|\eta_k|^2( \sum\limits_{j=1}^{K}(|\mathbf{h}^H_k(x)\mathbf{f}_j|^2)+|\mathbf{h}^H_k(x)\mathbf{f}_c|^2+\sigma_k^2)\geq \log(2)\sum\limits_{j=1}^{K}r_{c,j},\forall k.$}
\label{update_eq: common rate}
\end{align}   
\hrulefill
\end{figure*}
where $n_k \sim \mathcal{C}\mathcal{N}(0,\sigma_k^2)$ represents the receive noise for the $k^{th}$ user. According to the 1-layer RSMA scheme, the user decodes the common stream while regarding other streams as interference. After decoding the common stream and canceling it from the received signal, users proceed to decode their private streams. Thus, the signal-to-interference-plus-noise ratio (SINR) for common stream at the $k^{th}$ user can be expressed as
\begin{equation}
\text{SINR}_{c,k} = \frac{|\mathbf{h}_k^H(x)\mathbf{f}_c|^2}{\sum\limits_{j=1}^{K} |\mathbf{h}_k^H(x)\mathbf{f}_j|^2 + \sigma_k^2}.
\end{equation}
Consequently, the splitting common rate for the $k^{th}$ user is as follows
\begin{equation}
r_{c,k}=\log(1+\text{SINR}_{c,k}).\label{common communication Rate}
\end{equation}
The private SINR at the at the $k^{th}$ user is given by
\begin{equation}
    \text{SINR}_{p,k} = \frac{|\mathbf{h}_k^H(x)\mathbf{f}_k|^2}{\sum\limits_{j=1, j \neq k}^{K} |\mathbf{h}_k^H(x)\mathbf{f}_j|^2 + \sigma_k^2}.
\end{equation}
Accordingly, the private communication rate for the $k^{th}$ user is
\begin{equation}
r_{p,k}=\log(1+\text{SINR}_{p,k}).\label{private communication Rate}
\end{equation}

\subsection{Problem Formulation}
To verify the effectiveness of MAs in an RSMA system, we aim to maximize the sum of the private rate (\ref{private communication Rate}) and common rate (\ref{common communication Rate}) of all users. Hence, the optimization problem is formulated as
\begin{subequations}
\begin{align}
(\text{P1}) \max_{\mathbf{F}, \mathbf{r_c}, \mathbf{x}}  \mathcal{G}(\mathbf{F}, \mathbf{r_c}, \mathbf{x}) &= \sum\limits_{k=1}^{K} (r_{p,k}+r_{c,k}) \label{eq:originalG} \\
\quad \text{s.t.} \quad  \text{Tr}(\mathbf{F}^H \mathbf{F})& \leq P_0,  \label{eq: F}  \\
\quad  \log_2(1+\text{SINR}_{c,k})&\geq \sum\limits_{j=1}^{K} r_{c,j},\forall k\in\mathcal{K},\label{eq: common rate}\\
\quad  r_{c,k} &\geq 0,\forall k\in\mathcal{K},\label{eq: positive rate}\\
\quad X_{\min} &\leq x_i \leq X_{\max}, \forall i\in\mathcal{N},\label{eq:x1} \\
|x_i - x_j| &\geq D_0, \forall i,j \in\mathcal{N} ,i\neq j\label{eq:x2},
\end{align}
\end{subequations}
where $\mathbf{F}$ is the beamforming matrix and $P_0$ is the maximum transmit power at the BS. $\mathbf{r_c}=[r_{c,1},...r_{c,K}]^T$ and $r_{c,k}$ is the common rate allocated for the $k^{th}$ user. $D_0$ is the minimum separation distance between each pair of antennas to avoid the coupling effect \cite{MA_channel_model}. Constraint (\ref{eq: F}) is to ensure the power budget. Constraints (\ref{eq: common rate}) and (\ref{eq: positive rate}) refer to the common rate constraint. Constraints (\ref{eq:x1}) and (\ref{eq:x2}) define the feasible moving area of all MAs.

Solving (\text{P1}) presents significant challenges due to its nonconvex objective function and constraint (\ref{eq: common rate}) with respect to (w.r.t.) $\mathbf{F}$, $\mathbf{r_c}$, and $\mathbf{x}$. In the following section, efficient algorithms are developed to address this optimization problem.

\section{Proposed Solution}
In this section, we propose an AO algorithm to tackle the mutually coupled variables. Specifically, the beamforming matrix, splitting common rates, and antenna positions are updated in turn while the other parameters remain fixed until convergence is reached. (\text{P1}) can be divided into the following sub-problems.

\subsection{Updating Beamforming Matrix and Allocating Common Rates}
\begin{equation*}
    (\text{SP.1.1}) \max_{\mathbf{F},\mathbf{r_c}}{\mathcal{G}}(\mathbf{F},\mathbf{r_c}|\mathbf{x},\boldsymbol{\alpha},\boldsymbol{\beta})~\quad \\ \text{s.t.} \ \text{(\ref{eq: F}),(\ref{eq: positive rate}),(\ref{update_eq: common rate})}.
\end{equation*}

To address this non-convex subproblem, we employ the FP approach \cite{FPmethod,optimal_RSMA}. Specifically, we introduce auxiliary variables $\boldsymbol{\alpha}=[{\alpha}_1,{\alpha}_2,...,{\alpha}_K]^T$, $\boldsymbol{\beta}=[{\beta}_1,{\beta}_2,...,{\beta}_K]^T$, $\boldsymbol{\mu}=[\mu_1,\mu_2,...,\mu_{K}]^T$ and $\boldsymbol{\eta}^c=[{\eta}^c_1,{\eta}^c_2,...,{\eta}^c_K]^T$ to reformulate the objective function in (\text{P1}) and constraint (\ref{eq: common rate}) into equivalent convex forms as shown in (\ref{convexG}) and (\ref{update_eq: common rate}) respectively. This yields the following new convex problem.

\begin{equation*}
    (\text{SP.1.2}) \max_{\mathbf{F},\mathbf{r_c}}\hat{\mathcal{G}}(\mathbf{F},\mathbf{r_c}|\mathbf{x},\boldsymbol{\alpha},\boldsymbol{\beta})~\quad \\ \text{s.t.} \ \text{(\ref{eq: F}),(\ref{eq: positive rate}),(\ref{update_eq: common rate})}.
\end{equation*}

After transformation, the subproblem is convex. Hence, we can use the standard convex problem solvers such as CVX to solve the problem. The updating process of introduced auxiliary variables is discussed in the next subsection.

\subsection{Updating Auxiliary Variables for Quadratic Transform}
We can update the auxiliary variables with other parameters by fixing the following subproblems.

\begin{figure*}[t]
    \begin{equation}
        \frac{\partial \hat{\mathcal{G}}_{1,k}(x_n)}{\partial x_n} =  \sqrt{\alpha_k+1}  \operatorname{Re} \left\{ \sum_{l=1}^{L_p} j\rho_{k,l}\frac{2\pi}{\lambda}\cos{\theta_{k,l}e^{j\frac{2\pi}{\lambda}x_n\cos{\theta_{k,l}}}f_k^*\beta_k} \right\},
        \label{F1_xn}
    \end{equation}
    \begin{equation}
         \frac{\partial \hat{\mathcal{G}}_{2,k}(x_n)}{\partial x_n}=|\beta_k|^2  \operatorname{Re}\left\{\sum_{l=1}^{L_p}  \sum_{\substack{p=1}}^{L_p} \sum_{\substack{m=1}}^{N}  j\frac{2 \pi}{\lambda}\cos \theta_{k,l} \rho_{k,l}^* \rho_{k,p} f_{j,n} f_{j,m}^* e^{-j\frac{2\pi}{\lambda}x_n\cos{\theta_{k,l}}} e^{j\frac{2\pi}{\lambda}x_m\cos{\theta_{k,p}}} \right\}.
         \label{F2_xn}
    \end{equation}
    
\hrulefill
\end{figure*}

\begin{equation*}
    (\text{SP.2}) \max_{\boldsymbol{\alpha},\boldsymbol{\mu}}\hat{\mathcal{G}}(\boldsymbol{\alpha}|\mathbf{x}, \mathbf{F},\mathbf{r_c},\boldsymbol{\beta})~\quad \\ \text{s.t.} \ \text{(\ref{update_eq: common rate})},
\end{equation*}

\begin{equation*}
    (\text{SP.3}) \max_{\boldsymbol{\beta},\boldsymbol{\eta}}\hat{\mathcal{G}}(\boldsymbol{\beta}|\mathbf{x}, \mathbf{F},\mathbf{r_c},\boldsymbol{\alpha})~\quad \\ \text{s.t.} \ \text{(\ref{update_eq: common rate})}.
\end{equation*}

Given that the objective function w.r.t. $\boldsymbol{\alpha}$ and  $t_k(\mu_k|\mathbf{F},\eta_k)$ w.r.t. $\mu_k$ are both concave, we can update these variables by simply setting the partial derivatives to zero. Specifically, we set $\frac{\partial \hat{\mathcal{G}}(\boldsymbol{\alpha}|\mathbf{x}, \mathbf{F},\mathbf{r_c},\boldsymbol{\beta})}{\partial \boldsymbol{\alpha}} = 0 $ and $\frac{\partial t_k(\mu_k|\mathbf{F},\eta_k)}{\partial \mu_k} = 0 $. Since there is no constraint for $\boldsymbol{\alpha}$ and updated $\boldsymbol{\mu}$ will not influence the correctness of (\ref{update_eq: common rate}), we can directly derive the update formulas as follows.

\begin{equation}
\alpha_k^*=\frac{|\mathbf{h}_k^H(x)\mathbf{f}_k|^2}{\sum\limits_{j=1, j \neq k}^{K} |\mathbf{h}_k^H(x)\mathbf{f}_j|^2 + \sigma_k^2},
\label{eq: alpha}
\end{equation}

\begin{equation}
\mu_k^*=\frac{|\mathbf{h}_k^H(x)\mathbf{f}_c|^2}{\sum\limits_{j=1}^{K} |\mathbf{h}_k^H(x)\mathbf{f}_j|^2 + \sigma_k^2}.
\label{eq: mu}
\end{equation}

Similarly, the objective function is concave w.r.t. $\boldsymbol{\beta}$ and $t_k(\eta_k|\mathbf{F},\mu_k)$ is concave w.r.t. $\eta_k$. By setting the patio derivatives to zero, i.e., $\frac{\partial \hat{\mathcal{G}}(\boldsymbol{\beta}|\mathbf{x}, \mathbf{F},\mathbf{r_c},\boldsymbol{\alpha})}{\partial \boldsymbol{\beta}} = 0 $ and $\frac{\partial t_k(\eta_k|\mathbf{F},\mu_k)}{\partial \eta_k} = 0 $, i.e., 

\begin{equation}
\beta_k^*=\frac{\sqrt{1+\alpha_k}\mathbf{h}_k^H(x)\mathbf{f}_k}{\sum\limits_{j=1}^{K} |\mathbf{h}_k^H(x)\mathbf{f}_j|^2 +\sigma_k^2},
\label{eq: beta}
\end{equation}

\begin{equation}
\eta_k^*=\frac{\sqrt{1+\mu_k}\mathbf{h}_k^H(x)\mathbf{f}_c}{\sum\limits_{j=1}^{K} |\mathbf{h}_k^H(x)\mathbf{f}_j|^2 + |\mathbf{h}_k^H(x)\mathbf{f}_c|^2+\sigma_k^2}.
\label{eq: eta}
\end{equation}

Thus, the updating process of the beamforming matrix, splitting common rates and auxiliary variables, is summarized in \textbf{Algorithm \ref{updatingbeamforming}}.

 \begin{algorithm}[t]
    \renewcommand{\algorithmicrequire}{\textbf{Initialization:}}
	\renewcommand{\algorithmicensure}{\textbf{Output:}}
    \caption{FP for Solving Subproblems (\text{SP.1.2}), (\text{SP.2}) and (\text{SP.3}) .}
    \label{algo1}
    \begin{algorithmic}[1]

            \STATE Update $\mathbf{F}$ and $\mathbf{r_c}$ by solving problem (\text{SP.1}) with CVX;

            \STATE Update auxiliary variables $\alpha_k$ and $\mu_k$ by (\ref{eq: alpha}) and (\ref{eq: mu}) for $k\in \mathcal{K}$;

            \STATE Update auxiliary variables $\beta_k$ and $\eta_k$ by (\ref{eq: beta}) and (\ref{eq: eta}) for $k\in \mathcal{K}$;

        \ENSURE $\mathbf{F}$, $\mathbf{r_c}$, $\boldsymbol{\alpha}$, $\boldsymbol{\mu}$, $\boldsymbol{\beta}$, $\boldsymbol{\eta}$.
    \end{algorithmic}
    \label{updatingbeamforming}
\end{algorithm}

\subsection{Updating Antenna Positions}
With other parameters fixed, we need to tackle the subproblem as follows.
\begin{equation*}
    (\text{SP.4}) \max_{\mathbf{x}}\hat{\mathcal{G}}(\mathbf{x}|\mathbf{F},\mathbf{r_c},\boldsymbol{\alpha},\boldsymbol{\beta})~\quad \\ \text{s.t.} \ \text{(\ref{eq:x1}),(\ref{eq:x2}),(\ref{update_eq: common rate})}.
\end{equation*}

\begin{algorithm}[t]
    \renewcommand{\algorithmicrequire}{\textbf{Initialization:}}
	\renewcommand{\algorithmicensure}{\textbf{Output:}}
    \caption{AO for Solving Problem (\text{P1}) based on CFGS.}
     \label{algo2}
    \begin{algorithmic}[1]
        \REQUIRE Generate all the possible position alignments of the transmit antenna as $\{ \bm{\zeta}_{x1}, \bm{\zeta}_{x2}, \cdots, \bm{\zeta}_{xq_{x}} \}$ from $\mathcal{S}_{X}$. Set iteration index $l=1$ and sum rate $\text{SR}^{(0)}$ =0.
        \FOR{$i=1,2,\cdots, q_{x}$}

            \STATE Let $\mathbf{x}=\bm{\zeta}_{xi}$;
            \STATE Obtain $\mathbf{F}$, $\mathbf{r_c}$, $\boldsymbol{\alpha}$, $\boldsymbol{\mu}$, $\boldsymbol{\beta}$, $\boldsymbol{\eta}$ with \textbf{Algorithm \ref{algo1}};
            \STATE Compute $\text{SR}_i^{(0)}=\hat{\mathcal{G}}(\mathbf{F}, \mathbf{r_c}, \mathbf{x},\boldsymbol{\alpha},\boldsymbol{\beta})$;

        \ENDFOR 
        \STATE Let $\mathbf{x}^{(0)}={\bm{\zeta}}_{xk},  k=\arg\max$ $\text{SR}_k^{(0)}$;
        \REPEAT
            \REPEAT

            \STATE Updating $\mathbf{F}^{(l)}$, $\mathbf{r_c}^{(l)}$, $\boldsymbol{\alpha}^{(l)}$, $\boldsymbol{\mu}^{(l)}$, $\boldsymbol{\beta}^{(l)}$, $\boldsymbol{\eta}^{(l)}$ with \textbf{Algorithm \ref{algo1}} 
            \UNTIL{ the value of objective function converges}
            
            \STATE Adjust $\mathbf{x}^{(l)}$ as (\ref{eq: updatex});
            \WHILE{$\hat{\mathcal{G}}(\mathbf{F}^{(l)}, \mathbf{r_c}^{(l)}, \mathbf{x}^{(l)}, \boldsymbol{\alpha}^{(l)}, \boldsymbol{\beta}^{(l)})\leq\text{SR}^{(l-1)}$ or $\mathbf{x}^{(l)}\notin R_s$}
                \STATE $\kappa=\kappa/1.5$ and update $\mathbf{x}^{(l)}$ as (\ref{eq: updatex});
            \ENDWHILE   
            \STATE$\text{SR}^{(l)}=\hat{\mathcal{G}}(\mathbf{F}^{(l)}, \mathbf{r_c}^{(l)}, \mathbf{x}^{(l)}, \boldsymbol{\alpha}^{(l)}, \boldsymbol{\beta}^{(l)})$;
            \STATE Update iteration index $l=l+1$.
        \UNTIL{the value of the objective function converges.}
        \ENSURE $\text{SR}^{(l-1)}$.
    \end{algorithmic}
    \label{algoCFGS}
\end{algorithm}

Given the problem's non-convexity, obtaining optimal solutions is challenging. Drawing from the approach for updating antenna positions in \cite{Peng2024JointAP}, we adopt a two-stage CFGS method to obtain high-quality antenna positions.

In the first stage, discrete samples are taken at intervals of 1 $\lambda$, forming a point set denoted as $\mathcal{S}_{X}$. Samples comprising $N_T$ points combinations are obtained from $\mathcal{S}_{X}$. To reduce the complexity of the coarse search, the objective function values of each sample are evaluated after a single iteration of updating the beamforming matrix, rate splitting, and auxiliary variables, and the points from the sample that yield the highest performance are chosen as the initialization points for the subsequent stages. 

In the second stage, a gradient ascent method is employed to fine-tune the antenna positions. The gradient of (\ref{convexG}) w.r.t. the position of the $n^{th}$ antenna, $x_n$, is given by $\nabla \hat{\mathcal{G}}(x_n)=\frac{2}{\log(2)}\sum\limits_{k=1}^{K}(\frac{\partial \hat{\mathcal{G}}_{1,k}(x_n)}{\partial x_n}-\frac{\partial \hat{\mathcal{G}}_{2,k}(x_n)}{\partial x_n})^T$, where $\frac{\partial \hat{\mathcal{G}}_{1,k}(x_n)}{\partial x_n}$ and $\frac{\partial \hat{\mathcal{G}}_{2,k}(x_n)}{\partial x_n}$ are defined in (\ref{F1_xn}) and (\ref{F2_xn}). The update rule for the antenna positions is then given by
\begin{equation}
\mathbf{x}^{(l+1)}=\mathbf{x}^{(l)}+\kappa[\nabla \hat{\mathcal{G}}(x_1), \nabla \hat{\mathcal{G}}(x_2), \cdots, \nabla \hat{\mathcal{G}}(x_{N_T})]^T.
    \label{eq: updatex}
\end{equation}

However, applying gradient ascent directly may result in $\mathbf{x}$ moving out of the feasible region defined by constraints (\ref{update_eq: common rate}), (\ref{eq:x1}), (\ref{eq:x2}). To ensure that these constraints are consistently met throughout the optimization process, we adjust the step size in each iteration \cite{MeiGA}. Let $R_s$ denote the feasible region satisfying constraints (\ref{update_eq: common rate}), (\ref{eq:x1}), (\ref{eq:x2}). In the $(s+1)$-th iteration, if $\mathbf{x}^{(s+1)}$ is not within the feasible region, i.e., $\mathbf{x}^{(s+1)}\notin R_s$, the step size $\kappa$ is reduced to $\kappa /1.5$. This process repeats until $\mathbf{x}^{(s+1)}\in R_s$. Based on the subproblems discussed, the overall algorithm to solve (\(\text{P1}\)) is summarized in \textbf{Algorithm \ref{algoCFGS}}. 

\begin{figure}[t]
    \centering
    \includegraphics[width=1\linewidth]{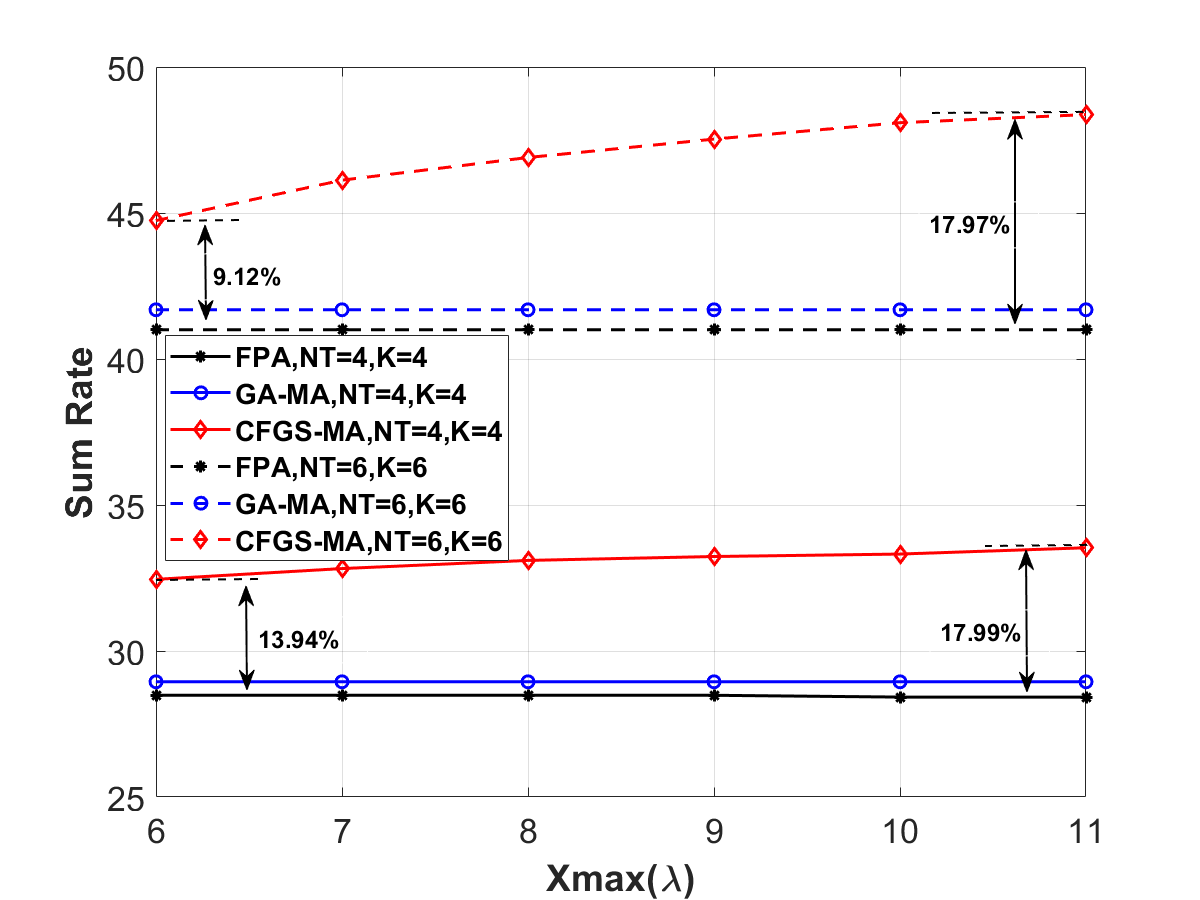}
    \caption{RSMA performance across different movable ranges, $P_0=30$ dBm.}
    \label{fig:range_comparison}
\end{figure}

\subsection{Complexity Analysis}
Since $\hat{\mathcal{G}}(\mathbf{F}, \mathbf{r_c}, \mathbf{x}, \boldsymbol{\alpha},\boldsymbol{\beta})$ is non-decreasing with each iteration and has an upper bound and updating $t_k(\mathbf{F},\mu_k,\eta_k)$ does not affect influence the validity of (\ref{update_eq: common rate}), \textbf{Algorithm \ref{algo2}} is guaranteed to converge. The computational complexity for conducting \textbf{Algorithm \ref{algo1}} once is approximately $\mathcal{O} \big((KN_T)^{3.5}\big)$ \cite{optimal_RSMA, FPmethod}. Consequently, the complexity for the coarse-grained search is $\mathcal{O} \big(I_c(KN_T)^{3.5}\big)$, where $I_c$ denotes the iteration number for the coarse-grained search. Since the complexity of fine-grained adjusting antenna positions mainly depends on (\ref{F2_xn}). Hence the complexity is approximately $\mathcal{O} \big(I_gL_p^2N_T^2K^2\big)$, where $I_g$ denotes the iteration number of gradient ascent. Assume $I_f$ denotes the iteration number required \textbf{Algorithm \ref{algo1}} to converge and $I_a$ represents the iteration number of the AO among the sub-problems. Therefore, the total complexity for \textbf{Algorithm \ref{algo2}} is $\mathcal{O} \bigg(I_c(KN_T)^{3.5}+I_a\big(I_gL_p^2N_T^2K^2+I_f(KN_T)^{3.5}\big)\bigg)$.

\section{Numerical Results}
In addition to the proposed CFGS-based algorithm with MAs \textbf{(CFGS-MA)}, we compare its performance with two other schemes: the fixed position antenna \textbf{(FPA)} scheme and the movable antenna scheme utilizing gradient ascent \textbf{(GA-MA)}. In the FPA scheme, antennas are positioned at a fixed spacing of $\lambda/2$, providing a static array configuration without adjustments. In contrast, in the GA-MA scheme, the antenna positions are assumed to be separated with $1~\lambda$ without the coarse-grained search. The initial beamforming matrix configuration follows \cite{SR_RSMA}. Specifically, the private beamforming vectors are designed using maximum ratio transmission (MRT). The common beamforming vector is constructed based on the dominant left singular vector of the channel matrix $\mathbf{H}$, where $\mathbf{H}=[\mathbf{h}_1,\cdots,\mathbf{h}_K]$.

In this study, we consider a path number of $L_p = 8$ and assume $K = 4$ users. Users are randomly positioned around the BS within an angular range of $[-\frac{\pi}{2},\frac{\pi}{2}]$. The distances between users and the BS are randomly distributed within $[20\,\text{m}, 100\,\text{m}]$. Consider a carrier frequency of 3 GHz, corresponding to a carrier
wavelength of $\lambda=0.1$ m. The elements of PRV follow a circularly symmetric complex Gaussian distribution (CSCG) distribution, i.e., $\rho_{k, i} \sim  \mathcal{CN}\left(0, \frac{C_{k}^2}{L_p}\right)$. The term $C_{k}^2=C_0D_k^{-\tau}$ represents the large-scale path loss, where $C_0=(\frac{\lambda}{4\pi^2})^2$ is the expected average channel power gain at the reference distance of 1 m. The path-loss exponent $\tau= 2.8$ is the path-loss exponent, and $D_k$ is the distance between the $k^{th}$ user and BS, respectively. Every user's received noise power is assumed to be $-90$ dBm. Additionally, without loss of generality, the feasible lower bound for MAs $X_{\min}$ is set to 0, and $D_0$ is set to $\frac{\lambda}{2}$. 

\begin{figure}[t]
    \centering
    \includegraphics[width=1\linewidth]{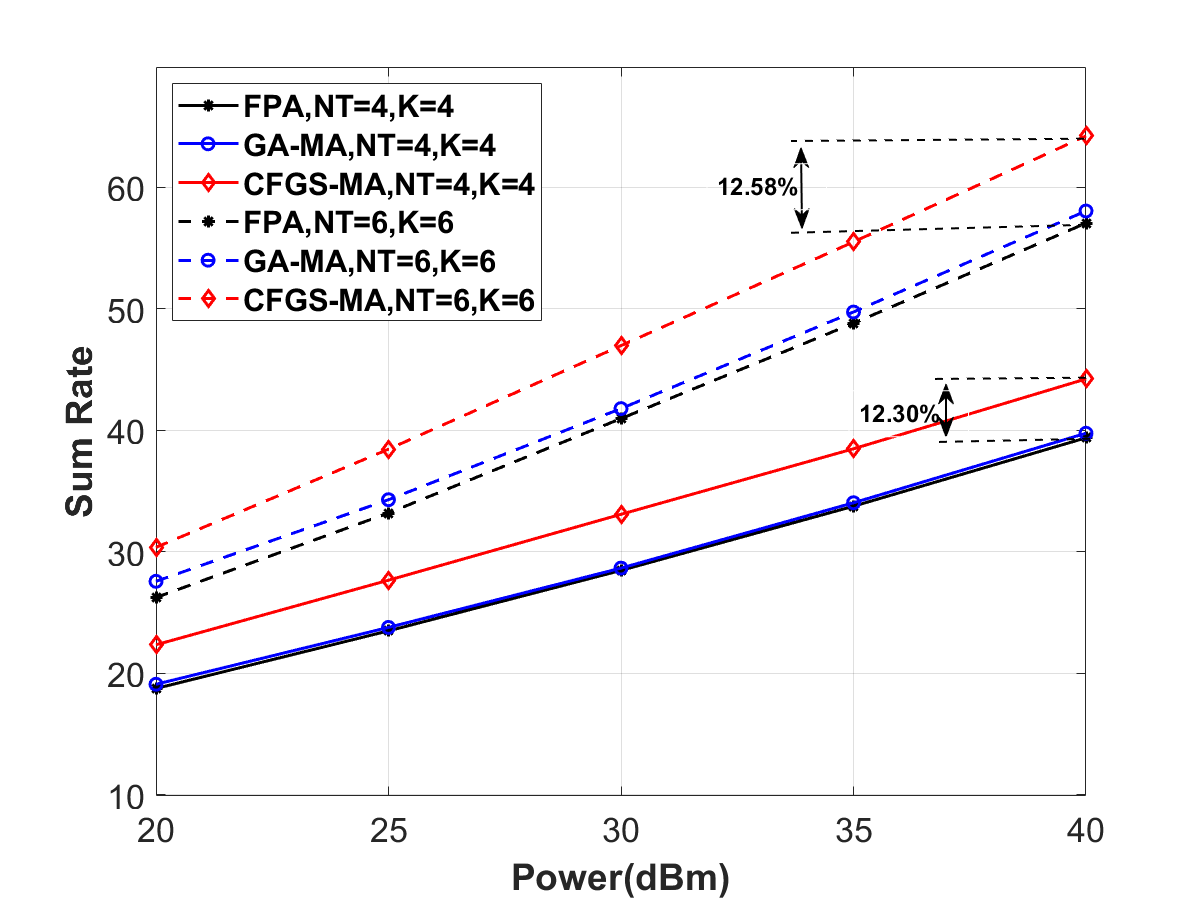}
    \caption{RSMA performance across different transmit powers, $X_{\max}=8\lambda$.}
    \label{fig:power_comparison}
\end{figure}

Fig. \ref{fig:range_comparison} illustrates the RSMA performance under varying movable ranges of BS antennas, as well as with different numbers of antennas and users. The range of movement for the transmit antennas spans from 6 $\lambda$ to 10 $\lambda$. We compare two configurations: one with 4 antennas serving 4 users and the other with 6 antennas serving 6 users. The results indicate that the gain achieved by the GA-MA scheme is relatively small, as it often converges to suboptimal positions, limiting its ability to exploit the increased movable range. In contrast, the CFGS scheme, with its initial coarse-grained search, demonstrates significant performance improvements. For the scenario with \( N_T = 4 \) and \( K = 4 \), the CFGS approach achieves up to a 17.99\% performance improvement over a range of 10 \( \lambda \), and a 13.94\% increase over a 6 \( \lambda \) range. Similarly, with \( N_T = 6 \) and \( K = 6 \), the CFGS method provides a 17.97\% gain over a 10 \( \lambda \) range and a 9.12\% improvement over a 6 \( \lambda \) range. These results indicate that as the number of antennas increases, the performance benefit from an extended movable range becomes more significant. Overall, the figure demonstrates that CFGS-MA can effectively leverage the expanded movable range to enhance performance across various antenna and user configurations.

Fig. \ref{fig:power_comparison} illustrates RSMA performance across varying transmit power levels with different antenna and user configurations. The transmit power ranges from 20 dBm to 40 dBm. The results reveal that the sum rate increases with higher power and as the number of antennas and users grows. Specifically, the GA-MA provides only a slight performance gain at a high transmit power level of 40 dBm. In contrast, the proposed CFGS-MA scheme achieves significant gains, delivering a $12.30\%$ gain in the scenario with \( K = 4 \) and \( N_T = 4 \), and a $12.58\%$ improvement with \( K = 6 \) and \( N_T = 6 \). This pattern persists across all tested transmit power levels and for each user and antenna configuration within the simulation, establishing that the CFGS-MA consistently outperforms both GA-MA and FPA. Such results underscore the robustness and efficiency of CFGS-MA in enhancing the sum rate compared to alternative schemes.

\section{Conclusion}
In conclusion, this work presented a novel MA-aided downlink RSMA system that, for the first time, fully leveraged spatial flexibility through dynamic adjustments of antenna positions at the base station. By jointly optimizing MA positions, the beamforming matrix, and common rate allocation, we aimed to maximize the system sum rate, encompassing both common and private messages. Although this led to a challenging non-convex optimization problem, we first used the AO method to decouple the parameters, subsequently addressing the problem by employing FP and a CFGS algorithm to obtain suboptimal solutions. The effectiveness of the proposed scheme was validated through simulations, which demonstrated that the use of MAs significantly improved both the performance and reliability of RSMA compared to traditional fixed-position antenna configurations. These results underscore the potential of MA-aided RSMA systems as a promising approach for future wireless communication networks, particularly in scenarios that demand high throughput and robust performance.



\bibliographystyle{IEEEbib}
\bibliography{IEEEabrv,myref}

\end{document}